\documentstyle[11pt 
]{article}  
\begin{document}  
\thispagestyle{empty}  
\rightline{UOSTP-01101} 
\rightline{KIAS-P01008}  
\rightline{{\tt hep-th/0102137}}  
  
\

\def\tr{{\rm tr}\,} \newcommand{\beq}{\begin{equation}}  
\newcommand{\eeq}{\end{equation}} \newcommand{\beqn}{\begin{eqnarray}}  
\newcommand{\eeqn}{\end{eqnarray}} \newcommand{\bde}{{\bf e}}  
\newcommand{\balpha}{{\mbox{\boldmath $\alpha$}}}  
\newcommand{\bsalpha}{{\mbox{\boldmath $\scriptstyle\alpha$}}}  
\newcommand{\betabf}{{\mbox{\boldmath $\beta$}}}  
\newcommand{\bgamma}{{\mbox{\boldmath $\gamma$}}}  
\newcommand{\bbeta}{{\mbox{\boldmath $\scriptstyle\beta$}}}  
\newcommand{\lambdabf}{{\mbox{\boldmath $\lambda$}}}  
\newcommand{\bphi}{{\mbox{\boldmath $\phi$}}}  
\newcommand{\bslambda}{{\mbox{\boldmath $\scriptstyle\lambda$}}}  
\newcommand{\ggg}{{\boldmath \gamma}} \newcommand{\ddd}{{\boldmath  
\delta}} \newcommand{\mmm}{{\boldmath \mu}}  
\newcommand{\nnn}{{\boldmath \nu}}  
\newcommand{\diag}{{\rm diag}}  
\newcommand{\bra}{\langle}  
\newcommand{\ket}{\rangle}  
\newcommand{\sn}{{\rm sn}}  
\newcommand{\cn}{{\rm cn}}  
\newcommand{\dn}{{\rm dn}}  
\newcommand{\tA}{{\tilde{A}}}  
\newcommand{\tphi}{{\tilde\phi}}  
\newcommand{\bpartial}{{\bar\partial}}  
\newcommand{\br}{{{\bf r}}}  
\newcommand{\bx}{{{\bf x}}}  
\newcommand{\bk}{{{\bf k}}}  
\newcommand{\bq}{{{\bf q}}}  
\newcommand{\bQ}{{{\bf Q}}}  
\newcommand{\bp}{{{\bf p}}}  
\newcommand{\bP}{{{\bf P}}}  
\newcommand{\thet}{{{\theta}}}  
\newcommand{\tauu}{{{\tau}}}  
\renewcommand{\thefootnote}{\fnsymbol{footnote}}  
\  
  
\vskip 0cm  
\centerline{ \Large  
\bf Noncommutative Chern-Simons Solitons  
}  
\vskip .2cm  
  
\vskip 1.2cm  
\centerline{ Dongsu Bak,${}^a\!\!$  
\footnote{Electronic Mail: dsbak@mach.uos.ac.kr}  
 Sung Ku Kim,${}^b$  
Kwang-Sup Soh${}^c$ and Jae Hyung Yee${}^d$  
}  
\vskip 7mm  
\centerline{Physics Department,   
University of Seoul, Seoul 130-743 Korea${}^a$}  
\vskip0.3cm  
\centerline{Physics Department, Ewha Women's University,  
Seoul 120-750 Korea${}^b$}  
\vskip0.3cm  
\centerline{Physics Department, Seoul National University,  
Seoul 151-742 Korea${}^c$}  
\vskip0.3cm  
\centerline{Institute of Physics and Applied Physics,   
Yonsei University,  
Seoul 120-749 Korea${}^d$}  
\vskip0.4cm  
\vskip 3mm  
  
\vskip 1.2cm  
  
  
\begin{quote}  
{
The Chern-Simons theories on a noncommutative plane, which is shown  
to be describing the quantum Hall liquid, is considered. We 
introduce matter fields fundamentally coupled to the noncommutative  
Chern-Simons field. Exploiting BPS equations for 
the nonrelativistic Chern-Simons theory, we find the 
exact solutions of  multi vortices that are closely packed 
and exponentially localized.  
We determine the position, the size  and the angular 
 momentum explicitly. We then construct solutions of
 two spatially separated vortex, and determine the 
moduli dependence of  
the size and the angular momentum.  
We also consider the relativistic Chern-Simons theory and find  
nontopological solutions whose properties are similar to the  
nonrelativistic counterpart. However, unlike the nonrelativistic  
case, there are two branches of solutions for a given magnetic  
field and they cease to exist below certain   
noncommutativity scale.  
}   
\end{quote}  
  

\newpage  
\section{Introduction} 

The solitons arising in the noncommutative theories have attracted
much attention recently\cite{Gopakumar}-\cite{Corley}.
These include
the noncommutative monopoles\cite{Hashimoto1,Gross1,Gross2}, 
the vortices\cite{Jatkar,Bak,Park}, 
the unstable lower dimensional
D-branes\cite{Aganagic,Harvey,Witten,Corley},  
and so on. As in the case of $U(2)$ noncommutative 
monopoles\cite{Hashimoto1,Gross2},
some of the solutions are smoothly connected to the commutative  
ones. There are noncommutative solitons such as $2+1$
dimensional  scalar noncommutative solitons\cite{Gopakumar},
which cease to exist 
in the  commutative limit. 
In the latter case, the characteristic properties
 of the noncommutative solitons  
largely differ from the ordinary solitons and have not 
been fully 
explored yet.

Various aspects of the ordinary Chern-Simons solitons are 
well known thus far\cite{Hong,Jackiw2,Dunne}. 
In this note, we like to explore the
solitons in the noncommutative Chern-Simons theories.
(Related discussions of noncommutative Chern-Simons 
theories have appeared in 
Refs.~\cite{Chu}-\cite{Jabbari}.)
It was shown recently  that the quantum Hall liquid can be  
described by the Chern-Simons theories on the noncommutative 
plane\cite{Susskind}. We shall add fundamental matter  
fields to the system,
which
may be viewed as describing  
the density fluctuations of the quantum Hall  
liquid. 

We first consider the nonrelativistic Chern-Simons 
theory on the noncommutative plane, which reduces to the  
Jackiw-Pi model in the commutative limit\cite{Jackiw2}. 
We solve the BPS equations and find  
the exact solutions of multi vortices,  which is  
localized exponentially, carrying integer flux  
$\Phi=m >0$.  
This is quite contrasted to the  
case of the commutative Jackiw-Pi solitons  
where the matter density falls 
off only in a certain power of the radius. To study 
the properties of the noncommutative solitons, we employ 
the covariant position operator\cite{Gross,Bak3} 
that describes the real space location 
 of the fluid particles\cite{Susskind}. With help of 
this operator, we are able to define gauge  
invariant position, size  
and angular momentum of the vortices. Hence it is extremely 
useful to explore the `local' properties  of 
generic profiles in spite of the fact that all the gauge invariant quantities 
in the noncommutative gauge theories more or less 
require integration over space\footnote{The Seiberg-Witten map  
is a way to find the local gauge invariant quantities\cite{Seiberg}.}.  
In particular,  the moduli size of these noncommutative Chern-Simons  
vortices is found to be $\sqrt{\theta\,(m\!-\!1)}$ and the total 
angular momentum $\pi\kappa\, m (m-2)$
where $\theta$  is the noncommutativity  
parameter.  
The inter-vortex distance is roughly order of $\sqrt{\theta}$ 
and cannot be squeezed further beyond the  distance. 
 This can be understood as revealing the area preserving  
nature of the quantum Hall liquid. 
Thus the simple exact solutions describe the closely 
packed $m$ vortices. 
 
We construct explicitly the two vortex solutions separated  
spatially. There are four real moduli parameters, which describe 
the positions of each vortex. We further identify the moduli  
dependence of the size of the two vortex configuration as well  
as the angular momentum.  Identifying the angular momentum 
at infinite  separation as an intrinsic one,  
the orbital part of  angular momentum is found to be  
$\pi\kappa m\, (m-1)$ in  case of the closely packed  
vortices.  
 
We then study the general solutions of the BPS equations within 
the rotationally symmetric ansatz. We show that the $\Phi=1$ solution  
above  
is unique within the ansatz. Except this $\Phi=1$ solution, 
we have proven that all the other solutions within the ansatz  
ought to carry the flux whose absolute value is greater than  
unity. Hence it seems  clear that the minimal flux in the theory 
is one for the solitonic sector. 
 
Next we consider the relativistic Chern-Simons 
model\cite{Hong} on the  
noncommutative plane. We construct the nontopological solutions, 
which are quite similar to the nonrelativistic 
noncommutative vortices. However, unlike the nonrelativistic model, 
there are two branches of the nontopological solutions for the same  
magnetic field and these solutions cease to exist when the  
noncommutativity parameter becomes small. We also identify the  
size and the angular momentum and discuss the separation of  
two vortices. 
 
In Section 2, we describe the nonrelativistic Chern-Simons model  
and derive the BPS equations. In Section 3, we present the closely packed  
multi vortex solutions and discuss their properties. 
The separations of two closely packed vortices are studied in Section 4  
by constructing explicitly the solutions depending on the moduli  
parameters. 
In Section 5, we prove that, within the rotationally symmetric ansatz,  
the flux is limited to $\Phi \ge 1$ or $\Phi <-1$.  In Section 6, 
we   derive the BPS equation for the relativistic 
Chern-Simons model and 
construct the nontopological vortex solutions. The last section 
comprises concluding remarks and discussions.  
 
{\bf\large  Note added}: The problem in this note   
was originally 
suggested by one of the authors\cite{Bak}. While we were  
investigating the subject, a related paper has  
appeared\cite{Lozano}. We found 
some overlaps  in  Section 5.

\section{Noncommutative Chern-Simons Theories}  
  
The noncommutative Chern-Simons  
theories are described by  the Lagrangian
\begin{equation}  
L_{\rm CS}=  
-{\kappa\over 2} \int d^2 x\,\, \epsilon^{\alpha\beta\gamma}   
\Bigl(A_\alpha * \partial_\beta A_\gamma -{2i\over 3}   
A_\alpha * A_\beta * A_\gamma  
\Bigr)\,,  
\label{lagcs}  
\end{equation}   
where the $*$-product is defined by  
\begin{equation}  
f(x)* g(x)\equiv \Bigl(e^{-i{\theta\over 2}\epsilon^{ij}\partial_i   
\partial'_j} f(x) g(x')\Bigr){\Big\vert}_{x=x'}\,.  
\label{star}  
\end{equation}  
Here and below  $\theta$ will be taken to be  positive 
without loss of generality.  
One may present the theory in terms of operators
on the Hilbert space defined by   
\begin{eqnarray}  
[\hat{x}\,,\,\hat{y}]=-i\theta   
\label{commutation}  
\end{eqnarray}  
where the $*$-product  
between functions becomes the ordinary product between the   
operators.   
For a given   
function   
\begin{eqnarray}  
f(x,y)=\int {d^2 k\over (2\pi)^2} \,  
\tilde{f}(k) e^{i(k_x x+k_y y)},   
\end{eqnarray}  
the corresponding operator can be found  
 by  
the Weyl-ordered form of  
\begin{eqnarray}  
\hat{f}(\hat{x},\hat{y})=\int  {d^2 k\over (2\pi)^2} \,  
\tilde{f}(k) e^{i(k_x \hat{x}+k_y \hat{y})}.   
\end{eqnarray}   
From the mapping, one may easily verify that
$\int d^2x \,f$ is replaced by   
$2\pi\theta \,\tr \hat{f}$ and $\partial_i f$  
corresponds to $-{i\over \theta}\epsilon_{ij}[ \hat{x}_j, \hat{f}]$.  
Let us introduce the creation and   
annihilation operators by  
$c^\dagger\equiv {1\over \sqrt{2\theta}}{(x+iy)}$ and by  
$c\equiv {1\over \sqrt{2\theta}}   
{(x-iy)}$, which satisfy $[c,c^\dagger]=1$.    
In order to represent arbitrary operators 
we  shall use the occupation number basis by $O= \sum  
O_{kl}|k\rangle\langle l|$  
with the number operator $
c^\dagger c$.

In terms of   
operator-valued fields, the action  can be equivalently written as  
\begin{equation}  
L_{\rm CS}=  
-{\pi\kappa\theta} \,\, \epsilon^{\alpha\beta\gamma}   
\tr \Bigl(A_\alpha  \partial_\beta A_\gamma -{2i\over 3}   
A_\alpha  A_\beta  A_\gamma  
\Bigr)\,,  
\label{lagcs1}  
\end{equation}   
where hats are dropped  for simplicity.   
We introduce the covariant  
position operator $X_i$,  
\begin{equation}  
X_i=x_i-\theta \epsilon_{ij} A_j\,.  
\label{covariant}  
\end{equation}   
These operators  then transform covariantly, i.e. $X_i'=U^{\dagger} X_i U$  
under the gauge transformation  
\begin{equation}  
A_\mu \ \ \rightarrow\ \  A'_\mu= iU^{\dagger}\partial_\mu U  
+U^{\dagger}A_\mu U\,,  
\label{gaugetr}  
\end{equation}   
with $U$ satisfying $UU^\dagger=U^\dagger U =I$.  
Written in terms of the covariant position  operator, 
the Lagrangian becomes  
\begin{equation}  
L_{\rm CS}=  
-{\pi\kappa\over \theta} \,\,   
\tr \Bigl( -\epsilon_{ij} X_i  (\dot{X}_j-i[A_0, X_j])   
+2\theta  
A_0\Bigr)\,,  
\label{lagcs2}  
\end{equation}   
where we dropped total derivative terms. As discussed in Ref.\cite{Susskind},  
the position $X_i$ describes positions of fluid particles labeled by the  
the comoving coordinates $x_i$. The action is   
invariant under the noncommutative   
gauge transformation 
whose continuum version corresponds to   
the area preserving diffeomorphism. Since $x_i$ is just labels of   
fluid particles, the theory describing fluid should be invariant  
under the relabeling as far as the $x_i$-space (labeling) density of   
fluid particles is preserved.  The above Lagrangian describes  
charged incompressible fluid in the low energies. 
The noncommutativity parameter  $\theta$ is related to the 
the comoving space density $\rho_0$ of the liquid by 
\begin{equation}  
\rho_0= {1\over 2\pi\theta}\,. 
\end{equation}   
Thus in some sense, $2\pi\theta$ is the minimal area for a  
fluid particle occupying in the comoving space $x_i$. 
The parameter $\kappa$ is related to the filling fraction by 
\begin{equation}  
\nu\equiv {2\pi\rho_0\over{\cal  B}}= {1\over 2\pi \kappa}\,, 
\end{equation}   
where ${\cal B}$ describes the external magnetic field 
applied to the quantum Hall liquid with the unit electric charge
set to unity. 
  
We shall first consider  a nonrelativistic matter field   
minimally coupled to the noncommutative  Chern-Simons term.   
The system is described by  
\begin{equation}  
L=L_{\rm CS} + 2\pi \theta \tr\Bigl(i D_0\phi\,\, \phi^\dagger  
-{1\over 2}D_i\phi (D_i\phi)^\dagger +  
{1\over 2|\kappa|}(\phi\phi^\dagger)^2  
\Bigr)\,,  
\label{lagmatter}  
\end{equation}   
where we use the fundamental coupling defined by  
$D_\mu \phi\equiv \partial_\mu \phi -i A_\mu \phi$.   
This corresponds to a noncommutative version of the Jackiw-Pi model.  
As in the case of the commutative version of Jackiw-Pi model,  
this theory allows the BPS bound. To show this, let us first note   
that the Hamiltonian is given by  
\begin{equation}  
E=2\pi \theta \tr\Bigl(  
{1\over 2}D_i\phi (D_i\phi)^\dagger   
-{1\over 2|\kappa|}(\phi\phi^\dagger)^2  
\Bigr)\,.  
\label{hamiltonian}  
\end{equation}   
This can be rewritten as  
\begin{equation}  
E=2\pi \theta \tr\Bigl(  
{1\over 2}(D_1\pm i D_2)\phi [(D_1\pm i D_2)\phi]^\dagger   
\pm {1\over 2}
\epsilon_{ij}D_i J^j   
+{1\over 2|\kappa|} (\pm|\kappa| B -\phi\phi^\dagger)\phi\phi^\dagger  
\Bigr)\,,  
\label{hamiltonian2}  
\end{equation}  
where $J^i={1\over 2i}  
(D_i\phi\, \phi^\dagger-\phi (D_i\phi)^\dagger)$.   
Noting the Gauss law,  
\begin{equation}  
\kappa B=\phi\phi^\dagger\,,  
\label{gauss}  
\end{equation}  
 one recognizes  that the theory allows the BPS bounds.  
When $\kappa >0$ the BPS equation becomes   
\begin{equation}  
(D_1 + i D_2)\phi=0  
\label{bps}  
\end{equation}     
while   
\begin{equation}  
(D_1 - i D_2)\phi=0  
\label{bps2}  
\end{equation}   
for $\kappa <0$.  
Since the commutation relation (\ref{commutation})  
breaks the parity symmetry, there is no obvious mapping between these two  
BPS equations. We find that the former allows a simple localized   
solutions. Hence we shall mainly focus on this case.  
   
Using the covariant position operator,  
 the former BPS equations can be written as  
\begin{equation}  
K^\dagger\phi-\phi c^\dagger =0,\ \ \    
1-[K,K^\dagger]={\theta\over\kappa}\phi\phi^\dagger  
\label{bps3}  
\end{equation}   
where we define $K\equiv (X_1-iX_2)/\sqrt{2\theta}$.  
These BPS equation do not involve $A_0$. However, once  
the BPS equations are solved, $A_0$ can be determined from   
the equation of motion,  
\begin{equation}  
\kappa \epsilon_{ij}F_{0j}= J_i\,.  
\label{a0}  
\end{equation}   
Together with the BPS equations, one finds  
\begin{equation}  
A_0=-{\phi\phi^\dagger\over 2|\kappa|}\,,  
\label{a01}  
\end{equation}   
which agrees with the expression of the  
commutative case except for the  specific  ordering  
of the field operators.

\section{Closely Packed Multi-Vortex Solutions }  
The simple $m$-vortex solutions we found are  
\begin{eqnarray}  
\phi=\sqrt{{m\kappa \over\theta}} |m-1\ket \bra 0|\,,\ \ \  
K=P_m c P_m+ S_m c S^\dagger_m\,,  
\end{eqnarray}   
where we define the shift operator $S_m$ and the projection   
operator respectively by  
\begin{eqnarray}  
S_m=\sum^\infty_{n=0}|n+m\ket \bra n|,\ \ \   
P_m= \sum^{m-1}_{a=0}|a\ket \bra a|\,.  
\end{eqnarray}   
The magnetic field of the solution takes the  form  
\begin{eqnarray}  
B={m\over \theta} |m\!-\!1\ket \bra m\!-\!1|\,.  
\label{bfield}  
\end{eqnarray}    
The flux carried by the vortex is $\Phi (\equiv \theta\, \tr B)=m$.  
Hence the solution describes $m$ vortex solutions. The matter  
density    
\begin{eqnarray}  
\rho\equiv \phi\phi^\dagger  
={m\kappa\over \theta} |m\!-\!1\ket \bra m\!-\!1|\,,  
\end{eqnarray}    
is well localized exponentially. This can be seen clearly  
if one maps the operator to an ordinary function. The density is   
mapped to   
\begin{eqnarray}  
\rho=   
{2m\kappa\over \theta} (-1)^{m-1} L_{m-1} \left({2r^2\over \theta}\right)  
e^{-{r^2\over\theta}}\,,  
\end{eqnarray}    
where $r^2= x_i x_i$ and
 $L_n(x)$ is the $n$-th order Laguerre polynomial. 
This is quite contrasted to the behaviors of the commutative version of   
the Jackiw-Pi solitons where the density falls off only by a certain  
power of the radius. The magnetic field  takes quite different   
form from that of the exact noncommutative solitons of the Abelian Higgs  
model where $B={1\over \theta } P_m$. Namely,   
in case of the Abelian Higgs model, the field with total flux $m$   
are evenly distributed over  from $|0\ket \bra 0|$   
to $|m-1\ket \bra m-1|$  
while  it is concentrated on one  
component in the case of the above Chern-Simons vortex.  
  
To study the properties of these solitons, let us first   
consider the   
size of soliton configuration. This can be measured   
in terms of the covariant position operator.  
The center position is defined as   
\begin{eqnarray}  
R^i= {\tr X^i \rho\over \tr \rho}\,.  
\end{eqnarray}   
This definition is gauge invariant whereas  
${\tr x^i \rho\over \tr \rho}$ does not lead to a gauge   
invariant quantity.  
Evaluating the center position of the above solution,  
one finds that $R^i$ is vanishing for all $m$.  
Translations of the vortex solutions give solutions   
describing the vortices located at the other  
position. Under a translation, the fields transform  
as   
\begin{eqnarray}  
\delta A_i=a^j \partial_j  A_i\,, \ \ \ 
\delta \phi=a^i \partial_i\phi\,.  
\end{eqnarray}    
We add here the gauge transformation in order to make   
the changes covariant\cite{Jackiw1978}. 
In particular, we add the transformation   
by the gauge parameter  
$\delta\Lambda=-a_i A_i$, whose form coincides with  
that of  
the ordinary gauge theory. Hence the translation is   
described by the transformation 
\begin{eqnarray}  
\tilde\delta A_i=a^j \epsilon_{ji} B\,,\ \ \
\tilde\delta \phi=a^i D_i\phi\,.     
\end{eqnarray}    
The finite form of translation constructed in this manner  
is in general  quite complicated although the corrections occurs only  
in $m\times m$ and $m\times \infty$ sectors of $K$ and $\phi$  
respectively.  
When $m=1$, the expressions of the translated vortex becomes  
particularly simple. It is   
\begin{eqnarray}  
\phi=\sqrt{{\kappa \over\theta}} |0\ket \bra {\bf a}| \,,\ \  \ 
K=  S_1 c S^\dagger_1 +\sqrt{2\theta}\,\, (a_1\!-\!ia_2)|0\ket \bra 0| \,,  
\end{eqnarray}   
where $|{\bf a}\ket$ denotes a coherent state defined by  
$|{\bf a}\ket\equiv e^{{i\over\theta}\epsilon_{ij}a_i x_j}|0\ket$.   
In fact one may generate solutions located at different position  
for a general $m$ with a choice of gauge parameter   
$\delta \Lambda=-{1\over\theta}a_i \epsilon_{ij}x_j$, which is gauge  
equivalent to the above construction. The solution reads  
\begin{eqnarray}  
\phi=\sqrt{{m\kappa \over\theta}} |m-1\ket \bra {\bf a}| \,,\ \  \ 
K=P_m c P_m+ S_m c S^\dagger_m+\sqrt{2\theta} (a_1\!-\! ia_2)P_m\,,  
\label{translated}  
\end{eqnarray}    
with the center position $R_i=a_i$.
One may also estimate the size $\Delta$ of the solitons by evaluating  
\begin{eqnarray}  
\Delta^2\equiv  {\tr (X^i-R^i)(X^i-R^i) \rho\over \tr \rho}\,.  
\end{eqnarray}   
One then finds that $\Delta^2=\theta (m-1)$ on   
the solutions (\ref{translated}). As we will see later on, the size  
here is minimal in some sense. Namely the vortex cannot be squeezed   
further. Thus the minimal distance between vortices are  
of order $\sqrt{\theta}$. Furthermore, the minimal  
area taken by $m$ vortices scales roughly  as $m\theta$.  
The behaviors are quite contrasted to the noncommutative  
vortices in the Abelian  
Higgs model\cite{Bak,Park} where any number of  the noncommutative  
 vortices can be at the same position  
at least classically. 
This seems to 
be closely related to the area preserving nature of the underlying 
quantum Hall liquid.

The system is invariant under the rotation where the rotation  
group is $SO(2)$. The parity is broken, which can be seen from the   
commutation relation (\ref{commutation}).  
Thus angular momentum may be   
defined by following   
the Noether construction of the conserved charges.  
When we use the ordinary function description with the $*$-product,  
the system is invariant under the rotation  
\begin{eqnarray}  
\delta A_i= -x_k \epsilon_{kj} \partial_j A_i-\epsilon_{ij}A_j\,,\ \ \ 
\delta \phi= -x_k \epsilon_{kj} \partial_j \phi\,.  
\label{rotation}  
\end{eqnarray}   
This can be rewritten, in terms of $*$-product,  as  
\begin{eqnarray}  
&&\delta A_i= -{1\over 2}(x_k * \epsilon_{kj} \partial_j A_i  
+ \partial_j A_i * x_k \epsilon_{kj})  
-\epsilon_{ij}A_j\nonumber\\  
&&\delta \phi= -{1\over 2}(x_k \epsilon_{kj} * \partial_j \phi  
+\partial_j \phi * x_k \epsilon_{kj}   
)\,,  
\label{rotation2}  
\end{eqnarray}   
where we have used the identity $2x_i f(x)=x_i * f(x) +f(x) * x_i$  
in the description of the ordinary functions.  Back to the operator   
formulation, the rotations read  
\begin{eqnarray}  
&&\delta X_i= -{i\over 2\theta}[x_kx_k, X_i]  
-\epsilon_{ij}X_j\nonumber\\  
&&\delta A_0 = -{i\over 2\theta}[x_k x_k, A_0]\nonumber\\  
&&\delta \phi=\,\, -{i\over 2\theta}\,[x_kx_k,\,\, \phi\,]\,.  
\label{rotation3}  
\end{eqnarray}   
We again add a gauge transformation   
$\delta \Lambda =-{1\over 2\theta}(X_k X_k-x_k x_k)$ to make   
the variation  
covariant\cite{Jackiw1978}.  
The resulting transformation of fields are  
\begin{eqnarray}  
&&\tilde\delta X = -{i\over 2\theta}[X_k X_k, X_i]  
-\epsilon_{ij}X_j\nonumber\\  
&&\tilde\delta A_0 = -{1\over 2\theta}D_0 (X_k X_k)\nonumber\\  
&&\tilde\delta \phi=\,\, -{i\over 2\theta}\,(X_kX_k\, \phi-\phi\, x_k x_k)\,.  
\label{rotation4}  
\end{eqnarray}   
Note that the variation $\tilde\delta\phi$ is covariant too here.   
The transformation of $X_i$ can be written in terms of the variation of   
$A_i$ as $\tilde\delta A_i={1\over 2}(X_i B+B X_i)$ that is reduced to  
the expression for the covariant variation in the   
commutative gauge theory.  
The construction of the corresponding Noether charge is straightforward. The   
resulting expression for the angular momentum  
is  
\begin{eqnarray}  
Q_J=2\pi\theta \tr\left( \epsilon_{ij} X_i J^j -  
{\theta\over 2} D_i\phi(D_i\phi)^\dagger\right) \,.  
\label{spin}  
\end{eqnarray}   
In the commutative limit in which the second term on the right side of  
the above equation is dropped out,  
this expression reduces to the angular momentum  
of the commutative theory in  Ref.\cite{Jackiw2}.  
Now let us evaluate the angular momentum using the solution. One finds  
\begin{eqnarray}  
Q_J=\pi\kappa\, m\,(m-2)\,. 
\label{angle}   
\end{eqnarray}   
The angular momentum
 scales as the vortex number squared unlike the case of   
ordinary Jackiw-Pi soliton where it scales linearly in $m\!-\!1$.
Rather this agrees to the angular momentum 
found in the ordinary relativistic model\cite{Kim}, which 
reads
\begin{eqnarray}  
Q'_J=\pi\kappa\, \Phi\,(2N -\Phi)\,,
\label{angle1}  
\end{eqnarray}   
with $N$ denoting the vorticity. Since  $N=m-1$
and $\Phi=m$ for our solutions, one finds the agreement. 
In the next section, we shall show that the orbital part alone is
just $\pi\kappa\, m\,(m-1)$ by studying the separation 
of vortices.   
  
As will be shown later, there is   
no further solution of $m=1$ within rotationally   
symmetric ansatz. Hence the $m=1$ case above is  
a  unique solution that is rotationally symmetric.    
Let us consider possible moduli of the $m=1$   
solution.   
In the commutative case, there are three  
moduli parameters for the one vortex solution.  
Two are locations of a vortex and the remaining  
is a scale parameter. All of these are related to    
symmetry transformations. Namely, the position is generated by the   
translation while an arbitrary size can be generated by a scale  
transformation that is also a symmetry of the model.  
In the noncommutative case, we have already found two   
moduli parameters representing the positions of the vortex.   
The question is what happens to the scale parameter.  
Due to the noncommutative scale on which the theory  
is explicitly dependent, the scale symmetry is no longer   
a symmetry of the model. But one may still generate  
solutions for a given solution by  
the scale transformation   
\begin{eqnarray}  
x_i\rightarrow  b x_i\,,\ \ \
t\rightarrow  b^2 t  
\end{eqnarray}     
together with  
$\theta \rightarrow  b^2\theta$,  
where the factor $b^2$ can be understood from the fact that  
the noncommutativity scale has a dimension of length squared.  
Unfortunately, the above $m$-vortex solutions are   
invariant under this scale transformation and no new   
solutions are generated\footnote{For example, $|0\ket\bra 0|$  
is mapped to $ 2 e^{-{r^2\over\theta}}$ and it is clear that it is   
invariant under the scale transformation together with the   
change of $\theta$.}.   
Hence it seems plausible that   
the $m=1$ solution has only two moduli   
parameters as the moduli space of one soliton is   
in general generated by a symmetry   
transformation.

\section{Separation of  Two Vortices}  
  
In order to study the separation of two vortices,  
we first note that the translations of the $m$-coincident  
vortices in the last section generate a nontrivial   
modification of fields only in 
$m\times m$ and $m\times \infty$ sectors of  
$K$ and $\phi$ respectively. Since the separation can be   
understood as a translation of each vortex, we take an   
ansatz where  
\begin{eqnarray}  
\phi=\sum^{m-1}_{a=0} |a\ket \bra \psi_a|\,, \ \ \ 
K=\sum^{m-1}_{a=0} \sum^{m-1}_{b=0}V_{ab} |a\ket \bra b|   
+S_m c S^\dagger_m\,,  
\end{eqnarray}   
where $|\psi_a\ket$ refers to a generic state in the Hilbert  
space defined by $[x,y]=-i\theta$. The BPS equation reduces  
to   
\begin{eqnarray}  
V^\dagger\phi-\phi c^\dagger =0,\ \ \    
P_m-[V,V^\dagger]={\theta\over\kappa}\phi\phi^\dagger\,.  
\label{bps23}  
\end{eqnarray}   
  Now let us consider $m=2$ case. We choose a gauge  
where $B$ is diagonal. The the most general form  
of $V^\dagger$ takes  
\begin{eqnarray}  
V^\dagger= q_0 P_2 +q_1 |0\ket \bra 1| +q_2 |1\ket \bra 0|\,,   
\end{eqnarray}   
where $q_0$, $q_1$ and $q_2$ are complex numbers.  
Note that under a transformation  
\begin{eqnarray}  
\phi\rightarrow \phi e^{c^\dagger q^*_0-c q_0}\,,\ \ \ \ 
 V^\dagger\rightarrow V^\dagger-q_0 P_2\,,  
\end{eqnarray}   
the BPS equation is invariant, which corresponds to  
a translation.  Hence one may  set $q_0=0$ without   
loss of generality. Solving the first BPS equation is  
now straightforward. For $\phi_{00}=\lambda_1$  
and $\phi_{10}=\lambda_2$, the general solution reads  
\begin{eqnarray}  
\bra \psi_1|=\lambda_1 \bra {\rm ch}|+\lambda_2 q_1 \bra {\rm sh}|   
\,, \ \ \  
\bra \psi_2|=\lambda_1 q_2 \bra {\rm sh}| +\lambda_2 \bra {\rm ch}|\,,  
\end{eqnarray}   
where we define  
\begin{eqnarray}  
&&\bra {\rm sh}|=  \bra 1 |+{q_1 q_2\over \sqrt{3!}} \bra 3 |  
+{(q_1 q_2)^2\over \sqrt{5!}} \bra 5 |+\cdots  
\nonumber\\  
&&\bra {\rm ch}|=  \bra 0 |+{q_1 q_2\over \sqrt{2!}} \bra 2 |  
+{(q_1 q_2)^2\over \sqrt{4!}} \bra 4 |+\cdots  \,.
\end{eqnarray}   
Since we take the gauge where the magnetic field is diagonal,  
$\bra \psi_1| \psi_2\ket$ ought to vanish, which can be seen  
from the second BPS equation. This   
implies either $\lambda_1$ or $\lambda_2$ are zero.   
We take $\lambda_1=0$ and $\lambda_2=\lambda$ using the residual   
gauge symmetry that interchanges $|0\ket$ and $|1\ket$.  
One may further choose $\lambda$ to be real and positive.  
Inserting the expressions to the second BPS equation,  
we obtain a set of algebraic equations  
\begin{eqnarray}  
&&B_{00}={1\over\theta}(1+|q_1|^2-|q_2|^2)={\lambda^2\over \kappa}  
\left|{q_1\over q_2}\right| \sinh |q_1q_2|   
\nonumber\\  
&&  
B_{11}={1\over\theta}(1-|q_1|^2+|q_2|^2)={\lambda^2\over \kappa}  
 \cosh |q_1q_2|\,.   
\end{eqnarray}   
Dividing the first equation by the second equation, we get  
\begin{eqnarray}  
{1+|q_1|^2-|q_2|^2\over 1-|q_1|^2+|q_2|^2}=  
\left|{q_1\over q_2}\right| \tanh |q_1q_2|\,,   
\end{eqnarray}   
from which one may find $|q_2|$ for an arbitrary   
$|q_1|\in [0,\infty)$.  
Once we get  $|q_2|$ for a given $|q_1|$, $\lambda$  
can be determined from any of the above equation.  
When $|q_1|=0$, one finds $|q_2|=1$, which corresponds  
to  
the closely packed two vortex solution described in  
the previous section.  
From the equation it is obvious that the maximum   
of $||q_1|^2-|q_2|^2|$  
equals one. Let us now show that the minimum  
of $|q_1|^2+|q_2|^2$ is also one. To show this,  
let us assume that $|q_1|^2+|q_2|^2 < 1$. Then from the   
left hand side of the above equation, one finds  
that   
\begin{eqnarray}  
{1+|q_1|^2-|q_2|^2\over 1-|q_1|^2+|q_2|^2}>  
{2|q_1|^2 \over 1-|q_1|^2+|q_2|^2} >  
{|q_1|^2 \over 1-|q_1|^2} > |q_1|^2   \,.
\end{eqnarray}   
Noting ${\tanh |q_1 q_2|\over |q_1q_2|}<1$, we find that   
the right hand side is smaller than $|q_1|^2$. Hence we   
get a contradiction. Similarly, one may easily show that  
$|q_1|< |q_2|$ on the solution for all $|q_1|$.  
Furthermore, when $|q_1|$ become large,  
the difference between   
$|q_2|$ and $|q_1|$ becomes exponentially small  
as $\sim {e^{-2|q_1|^2}\over 2|q_1|}$.  
  
The overall position $R_i$ of this solution is zero.  
The size of the configuration is  
\begin{eqnarray}  
\Delta^2=\theta (|q_1|^2+|q_2|^2)\ge \theta\,.  
\end{eqnarray}   
The last inequality follows from the fact that the minimum  
of $(|q_1|^2+|q_2|^2)$ is one as shown above.  
We further compute the higher moments,   
\begin{eqnarray}  
 {\tr [(X^i-R^i)(X^i-R^i)]^n \rho\over \tr \rho}  
=[\theta (|q_1|^2+|q_2|^2)]^n\,.  
\end{eqnarray}   
In the commutative theory,  
this kind of behaviors for the higher moments  
can  be found  
only when two pointlike masses of an equal mass  
 are separated by a distance   
\begin{eqnarray}  
d=2\sqrt{\theta(|q_1|^2+|q_2|^2)}\,.  
\end{eqnarray}  
Hence we conclude that the configuration describes two separated  
 vortices at a distance $d$.  
The angular momentum of the configuration   
can be found as   
\begin{eqnarray}  
Q_J=2\pi\kappa \left( (|q_1|^2-|q_2|^2)^2-1\right)\,.  
\end{eqnarray}  
The maximum of the angular momentum occurs at   
the closely packed   
solutions in the previous section. The angular momentum decreases  
as the separation gets larger and becomes twice of the   
angular momentum of one vortex of $\Phi=1$ in the large   
separation limit. Therefore one may consistently regard each   
vortex  carrying an intrinsic angular momentum $-\pi\kappa$.   
The orbital part of the angular momentum in the two vortex case   
is  
\begin{eqnarray}  
Q_{\rm orb}=Q_J +m\pi\kappa=2\pi\kappa(|q_1|^2-|q_2|^2)^2\,.  
\end{eqnarray}  
Thus as claimed before,
 the orbital part of angular momentum   
 becomes $
\pi\kappa\, m\, (m-1)$ for the closely packed $m$-vortices.  
  
Finally let us count moduli parameters of these separated vortices.  
There are two real parameters describing the center position specified by   
a complex number $q_0$. $|q_2|$ is fixed by $|q_1|$ and  
the real parameter $\lambda$ is completely  
fixed in terms of $|q_1|$. Among the argument of $q_1$ and   
$q_2$, 
the relative part $(\varphi_{1}\!-\!\varphi_{2})$  can be gauged away   
and only overall part is gauge invariant. In fact, 
under the  rotation defined  
in (\ref{rotation4}), $q_1$ and $q_2$ change as $q'_1=q_1 e^{i\varphi}$   
$q'_2=q_2 e^{i\varphi}$. Thus we conclude that the overall argument   
describes the angular coordinates in the relative space of the two   
vortices. In total we have four real moduli parameters which   
specify the planar locations of two vortices.

\section{General Rotationally Symmetric Vortices}  
The rotationally symmetric configuration of the noncommutative   
Jackiw-Pi model was studied   
numerically in Ref.\cite{Lozano}.  
Nonetheless we repeat here the analysis to show certain analytic   
properties.  
We first consider the BPS equation with $\kappa >0$.  
We take  a generic rotationally symmetric  ansatz     
\begin{eqnarray}  
 \phi =\sum^\infty_{n=0}\phi_n|n+m-1\ket \bra n|\,, \ \ \ 
K^\dagger =\sum^\infty_{n=0}k^*_n|n+1\ket \bra n|\,,  
\end{eqnarray}  
for some positive integer $m$. One may easily verify that  
the other choice   $\phi =\sum^\infty_{n=0}\phi_n|n\ket \bra n+m|$  
does not lead to any solutions. The BPS equations then imply  
that  
\begin{eqnarray}  
&& \phi_{n+1}\sqrt{n+1}=\phi_n k^*_{n+m-1}\,,   
\nonumber\\  
&&B_{n+m-1}=  
{1\over\theta} (1-|k_{n+m-1}|^2+|k_{n+m-2}|^2)={1\over\kappa}  
 |\phi_{n}|^2\,,  
\label{recurrence}  
\end{eqnarray}  
for $n\ge 0$ with   
\begin{eqnarray}  
|k_a|=a+1, \ \ B_a=0 \ \ {\rm for}\  a< m-1\,.   
\end{eqnarray}  
Defining $|k_{n+m-1}|=n+v_n\,\,(\ge 0)$,   
one gets a recurrence relation  
\begin{eqnarray}  
v_{n+1}-v_n={n+v_n\over n+1}(v_n-v_{n-1})\,,   
\label{recurrence1}  
\end{eqnarray}  
for $n\ge 1$. The initial conditions are identified from  
(\ref{recurrence}) as  
 $ 0\le v_0 \le m$ and $v_1-v_0= -v_0(m-v_0)$.  
Since  $v_1-v_0\le 0$, $v_{n}-v_{n+1}$ is all   
nonnegative definite. Hence the series $v_n$ is either  
constant or monotonically decreasing.  
The total flux here is given by $\Phi=m-v_\infty$.  
  
There are two special points where $v_n$ is constant.  
When $v_0=m$, $v_n=m$ and $\phi_n=0$ for all $n$.  
The flux is zero and this corresponds to  the   
trivial vacuum   
solution.  
When $v_0=0$, $v_n=0$ and $|\phi_n|^2$ is nonvanishing  
only for $n=m-1$ with a value   
${m\kappa\over\theta}$. This is the closely packed solution  
we discussed in the previous section.  
Except these two cases, $v_\infty$ must be negative   
definite. To show this, let us assume $v_\infty\ge 0$.  
Since $v_n$ is monotonically decreasing, $v_n-v_{n+1}$ is then   
positive definite. From the recurrence relation   
(\ref{recurrence1}), we have  
\begin{eqnarray}  
v_{n}-v_{n+1} > {n\over n+1}(v_{n-1}-v_n)> \cdots > {1\over n+1}  
(v_0-v_1)> 0   
\end{eqnarray}  
for $n\ge 1$.  
Summing this for $l=1$ to $n$, one finds  
\begin{eqnarray}  
v_{1}-v_{n+1} > (v_0-v_1)\sum^n_{l=1} {1\over l+1}  
\,.  
\end{eqnarray}     
For large $n$, the right hand side diverges as $\ln n$ and  
we get a contradiction. Hence we get a bound for the total   
flux as  
\begin{eqnarray}  
\Phi=m-v_\infty > m  
\end{eqnarray}   
when $v_0\neq 0,m$.   
This proves the claim that the closely packed $\Phi=1$ solution is   
unique within the rotationally symmetric ansatz.  
The series for $0< v_0 <m$ case in general   
converges. Whether the flux is quantized or not for  
the generic initial data seems not clear though in   
Ref.~\cite{Lozano} it is claimed that they found   
that it is not quantized numerically. In case of the   
commutative Jackiw-Pi model, such solutions carrying   
noninteger flux  
 are excluded  
by the analyticity requirement although the solutions   
satisfy locally the equations of motion.  
We do not know how to implement such analyticity  
in the noncommutative case. In this respect,   
the quantization  
of the flux numbers are not quite clear. We leave this   
issue for the future study.  
  
Now let us turn to the case of $\kappa <0$.  
In this case an appropriate ansatz will be    
\begin{eqnarray}  
\phi =\sum^\infty_{n=0}\phi_n|n\ket \bra n+m-1|\,,\ \ \   
K^\dagger =\sum^\infty_{n=0}k^*_n|n+1\ket \bra n|\,.  
\end{eqnarray}  
The other choice   $\phi =\sum^\infty_{n=0}\phi_n|n+m\ket \bra n|$  
again does not lead to any solutions.  
Inserting the ansatz to the BPS equations for $\kappa <0$,
one gets the recurrence relation
\begin{eqnarray}  
&& \phi_{n+1}k_n=\phi_n  \sqrt{n+m}   
\nonumber\\  
&&B_{n}=  
{1\over\theta} (1-|k_{n}|^2+|k_{n-1}|^2)=-{1\over|\kappa|}  
 |\phi_{n}|^2  
\label{recurrence5}  
\end{eqnarray}  
for $n\ge 0$ with $k_{-1}=0$. Eliminating $\phi_n$ again, we get  
\begin{eqnarray}  
v_{n+1}-v_n={n+m\over n+v_n}(v_n-v_{n-1})\,,   
\label{recurrence12}  
\end{eqnarray}  
where we define again $|k_n|^2=n+v_n\ge 0$. The total   
flux is then given by  
$\Phi=1-v_\infty$.  
The initial conditions read $v_0\ge 1$ and   
$v_1-v_0= m(v_0-1)/v_0 \ge 0$. When $v_0=1$, we get the trivial   
vacuum solution again, so    
we shall consider only the case of  $v_0 >1$. $v_n$ is   
monotonically increasing and one may easily show that  
$v_\infty >m+1$ by a similar way adopted for $\kappa >0$.  
Hence we conclude that any nontrivial vortex solutions carry  
the flux  $\Phi < -1$.    
  
In summary we have proved that,  
within the rotationally symmetric ansatz,   
the $\Phi=1$ solution is unique   
and any other solutions should necessarily carry the flux with   
 $|\Phi|>1$.

\section{Nontopological Vortices in the Relativistic  
Chern-Simons Model}  
In this section we shall consider the relativistic Chern-Simons  
model with fundamentally coupled matter described by  
\begin{equation}  
L=L_{\rm CS} + 2\pi \theta \tr\Bigl( D_0\phi(D_0\phi)^\dagger  
-D_i\phi (D_i\phi)^\dagger - V(\phi\phi^\dagger)  
\Bigr)\,,  
\label{lagrelativistic}  
\end{equation}   
where the potential is  
\begin{equation}  
V(\xi)={1\over \kappa^2}\xi (v^2-\xi)^2\,.  
\label{potential}  
\end{equation}   
We shall derive here the BPS equation similarly to the   
case of the nonrelativistic model.  
The derivation was done in Ref.\cite{Lozano}, but we find it   
incomplete because they assume certain relation that   
is unnecessary.  
To begin with, let us first note that the Hamiltonian is  
given by  
\begin{equation}  
E=2\pi \theta \tr\Bigl(D_0\phi (D_0\phi)^\dagger   
+D_i\phi (D_i\phi)^\dagger +V(\phi\phi^\dagger)  
\Bigr)\,.  
\label{hamiltonianrel}  
\end{equation}  
We simply note that the above Hamiltonian  
can be rewritten identically as  
\begin{equation}  
E\!=\!2\pi \theta \tr\Bigl(|D_0\phi \pm {i\over \kappa}   
(v^2\!-\!\phi\phi^\dagger)\phi|^2   
+|(D_1\!\pm\! i D_2)\phi|^2  
\pm (\phi\phi^\dagger\!-\!v^2)(B-{j_0\over\kappa})  
\mp {\epsilon_{mn}\over 2} D_m j_n\pm v^2 B  
\Bigr)\,,  
\label{hamiltonianrel1}  
\end{equation}  
where we define the current $j_\mu$ as  
\begin{equation}  
j_\mu= i (D_\mu \phi\, \phi^\dagger- 
\phi (D_\mu\phi)^\dagger)\,,  
\end{equation}  
and $|O|^2\equiv O O^\dagger$ for any operator $O$.  
Now using the Gauss law constraint,  
\begin{equation}  
\kappa B=j_0 \,,  
\end{equation}  
and ignoring the total derivative term,  
we find   
\begin{equation}  
E=2\pi \theta \tr\Bigl(|D_0\phi \pm {i\over \kappa}   
(v^2\!-\!\phi\phi^\dagger)\phi|^2   
+|(D_1\!\pm\! i D_2)\phi|^2    
\pm v^2 B \Bigr)\ge  v^2 |\Phi|\,.  
\label{hamiltonianrel2}  
\end{equation}  
The saturation of the bound occurs if the BPS equations  
\begin{eqnarray}  
&&(D_1\!\pm\! i D_2)\phi=0\,,\nonumber\\  
&&D_0\phi \pm {i\over \kappa}   
(v^2\!-\!\phi\phi^\dagger)\phi=0  
\end{eqnarray}  
hold. Using again the Gauss law, the second equation  
can be rewritten as  
\begin{eqnarray}  
B=\pm {2\over \kappa^2}\phi\phi^\dagger (v^2-\phi\phi^\dagger)\,.  
\end{eqnarray}   

We shall not explore the full details of the soliton solutions 
present 
in this model. Instead, we shall focus on the nontopological soliton  
solutions. 
  For the closely packed solutions, one has two types of solutions 
for the same magnetic field: They are   
\begin{eqnarray}  
\phi=\lambda_\pm |m-1\ket \bra 0| 
\,,\ \ \  K=P_m c P_m+ S_m c S^\dagger_m\,,  
\end{eqnarray}   
where $\lambda_\pm$ is given by 
\begin{eqnarray}  
\lambda^2_\pm={v^2\over 2}\left( 
1\pm \left(1-{2m\kappa\over\theta v^4}\right)^{1\over 2} 
\right)\,.  
\end{eqnarray}  
The solution of this type exists only for $\theta  \ge  2m\kappa/v^4$. 
Hence it is clear that the solution do not exist in the commutative limit  
where $\theta$ 
goes to zero. The flux is again $\Phi=m$ and the energy is $2\pi m\,v^2 $ 
that of course saturates the BPS bound. 
The charge carried by the solution is  
$Q=2\pi\theta \tr j_0=2\pi\kappa \, m$. 
The position is defined as 
\begin{eqnarray}  
R_i\equiv {\tr X_i\, j_0 \over \tr j_0}\,, 
\end{eqnarray} 
and the center position vanishes on the solution. 
The size defined with respect to  
the charge density is $\Delta^2=\theta (m-1)$. 
 
The position can be defined with respect 
to the Hamiltonian 
density ${\cal H}$\footnote{We define here the Hamiltonian density 
operator as the covariant quantity inside the trace 
of (\ref{hamiltonianrel}).}: 
\begin{eqnarray}  
R^i_{H}\equiv {\tr X^i\, {\cal H} \over \tr {\cal H}}\,, 
\end{eqnarray} 
which  vanishes on the solution. 
One may also consider the matter distribution size as  
\begin{eqnarray}  
\Delta^2_{H}\equiv {\tr (X^i-R^i_{H}) (X^i-R^i_{H})\, {\cal H}  
\over \tr {\cal H}}\,. 
\end{eqnarray} 
For the vortex configuration, this size of the energy distribution 
is found to be 
\begin{eqnarray}  
\Delta^2_{H}= (m-1)\theta \left(1+{(m-2)\lambda^2_\pm\over m v^2}\right) \,. 
\end{eqnarray} 
The value is again zero for $m=1$ and there is no difference
between the two branches 
for $m=2$. For $m>2$,  the $(-)$ branch 
solution has more closely packed; Namely 
the minimal inter-vortex distance 
is  smaller in the $(-)$ branch. 
 
The angular momentum can be found in a similar way to the nonrelativistic  
case. Under the rotation  in (\ref{rotation4}), the Noether procedure  
leads to the angular momentum 
\begin{eqnarray}  
Q_J=2\pi\theta \tr\left( \epsilon_{ij} X_i T^{0j} -  
{i\theta\over 2} \Bigl( 
D_iD_0\phi (D_i\phi)^\dagger- 
D_i\phi(D_iD_0\phi)^\dagger 
\Bigr) 
\right) \,, 
\label{spinrel}  
\end{eqnarray}   
where $T^{0i}$ is the momentum density 
\begin{eqnarray}  
T^{0i}=-{1\over 2}\left(   
D_i\phi (D_0\phi)^\dagger  
+D_0\phi (D_i\phi)^\dagger  
\right) \,. 
\label{momemtum}  
\end{eqnarray}   
The expression again reduces to that of the ordinary field theory 
in the commutative limit.  
The angular momentum evaluated on the above solution is  
$\pi\kappa \,m\,(m-2)$.  
Hence the solutions describe the closely packed $m$ vortices and their  
properties are quite similar to those of nonrelativistic  
counterpart.  
One thing that is distinct from the  
nonrelativistic closely packed solution 
lies in the fact that there are two types of 
solutions for the same  given magnetic 
field configuration. 
In the $(-)$ branch solution, nonvanishing components of 
$(\phi\phi^\dagger)_{nn}$  is centered  around the symmetric  vacuum 
while  the components of $(+)$ branch solution  
 are centered around the broken vacuum. As $\theta$ grows, the values 
of $\lambda_\pm$ approaches the vacuum values.  
When $\theta v^4 = 2 m \kappa$, the two branches coincide and become  
just one solution.  
 
The separations of two closely packed solution can be separated again. 
The solution takes a form 
\begin{eqnarray}  
\phi= \lambda\, \Bigl(q_1 |0 \ket \bra  {\rm sh} | 
+|1 \ket \bra  {\rm ch} | 
\Bigr)\,,\ \ \   
K^\dagger= q_1 |0\ket \bra 1| +q_2 |1\ket \bra 0|   
+S_m c^\dagger S^\dagger_m\,.  
\end{eqnarray} 
The diagonal part of the $2\times 2$ sector of  $K^\dagger$ again 
describes the center position and we set it to zero using the  
translational symmetry.  
The parameters satisfy the algebraic equations   
\begin{eqnarray}  
&&B_{00}={1\over\theta}(1+|q_1|^2-|q_2|^2)={2\lambda^2\over \kappa^2}  
\left|{q_1\over q_2}\right| \sinh |q_1q_2|   
\left(v^2 - \lambda^2 \left|{q_1\over q_2}\right| \sinh |q_1q_2| \right)  
\nonumber\\  
&&  
B_{11}={1\over\theta}(1-|q_1|^2+|q_2|^2)={2\lambda^2\over \kappa^2}  
 \cosh |q_1q_2| \,\Bigl(v^2 - \lambda^2 \cosh |q_1q_2| \Bigr)\,.  
\end{eqnarray}   
The number of parameters involved are again four real parameters. 
The center position has two real parameters and $|q_1|$ 
and the overall angle  
of $q_1$ and $q_2$ are the remaining moduli. 
The relative angle can be gauged away  
and $|q_2|$ and $\lambda$ are determined in terms of $|q_1|$ by the  
above set of the equations.

\section{Conclusion} 
In this note we have  constructed the solutions of 
 the noncommutative Chern-Simons 
solitons and investigated their properties by  
computing the position, the size and the angular  
momentum. The nature of the 
noncommutative vortices
differ from those in the 
commutative version.

In case of the closely packed multi  
vortex solutions, we have considered  
their moduli separations only for the case of two  
vortices. For the general closely packed vortices with  
$m>2$, the ansatz taken in this note leads to a finite  
set of closed algebraic equations. Though there  
might arise a little complication, it is worth studying  
the detailed  
structure of their moduli dependence.  
In particular how the minimal size  
$m$-vortex configurations is achieved in a  
moduli  
dependent manner would be  
quite interesting. 
 On the nonrelativistic model, one problem unresolved is  
whether there is a quantization of flux generically. 
In Ref.\cite{Lozano}, it is suggested that 
the flux quantization of the vortex solutions does not occur 
if one just require the convergence of the difference  
equation. In case of the commutative Jackiw-Pi model, 
only the regularity requirement of the solution  
at the origin leads to a quantization. We do not know how to  
understand such conditions in the noncommutative case. Further  
study is required on the issue. 

The other issue we omitted in this note is on the 
low energy moduli dynamics. In the commutative 
Jackiw-Pi model, it is shown that the 
solitons behave as dual object of the elementary 
particle excitations\cite{Bak6}. How the dynamics occurs
in case of the noncommutative vortices need to be 
clarified. Also In this respect, the 
quantum nature of elementary excitation would be of 
interest\cite{Bak7}.  
 
In case of the relativistic model, we have not explored the 
solitons in the topological sector. The BPS equations suggest that, 
the solitons in this sector might have some common properties 
with the Abelian Higgs model since they involve a broken  
vacuum. There are actually models where the BPS equations are 
precisely the same as that of the Abelian-Higgs 
model\cite{Jatkar,Bak,Park}. 
They are the nonrelativistic models with repulsive  
interactions with the background charge density\cite{Barashenkov} 
or the  
external magnetic field\cite{Zhang}. 
Since some of the properties of solutions  
are known in this  case, the study of the low energy dynamics would be 
quite interesting.

\noindent{\large\bf Acknowledgment}   
DB would like to thank Kimyeong Lee and Jeong-Hyuck Park for  
enlightening discussions.   
This work is supported in part by KOSEF 1998  
Interdisciplinary Research Grant 98-07-02-07-01-5 (DB), by  
UOS Academic Research Grant (DB), by BK21 Project of 
Ministry of Education(KSS),
 and  by
 Korea Research Foundation under Project Nos. 99-005-D00009 (JHY)
and 99-015-DI0021 (SKK).

\vfill
\eject



\begin{thebibliography}{99}  
  


\bibitem{Gopakumar} 
R.~Gopakumar, S.~Minwalla and A.~Strominger, 
JHEP{\bf 0005}, 020 (2000) 
[hep-th/0003160]. 
 

\bibitem{Hashimoto1}
A.~Hashimoto and K.~Hashimoto,
JHEP{\bf 9911}, 005 (1999)
[hep-th/9909202];
D.~Bak, 
Phys.\ Lett.\ B {\bf 471}, 149 (1999) 
[hep-th/9910135]; 
K.~Hashimoto, H.~Hata and S.~Moriyama,
JHEP{\bf 9912}, 021 (1999)
[hep-th/9910196].


\bibitem{Gross1} 
D.~J.~Gross and N.~A.~Nekrasov, 
JHEP{\bf 0007}, 034 (2000) 
[hep-th/0005204];
D.~J.~Gross and N.~A.~Nekrasov, 
JHEP{\bf 0010}, 021 (2000) 
[hep-th/0007204]. 
 
 



\bibitem{Polychronakos} 
A.~P.~Polychronakos, 
Phys.\ Lett.\ B {\bf 495}, 407 (2000) 
[hep-th/0007043]. 




 
\bibitem{Jatkar} 
D.~P.~Jatkar, G.~Mandal and S.~R.~Wadia, 
JHEP{\bf 0009}, 018 (2000) 
[hep-th/0007078]. 
 
\bibitem{Bak5} 
D.~Bak and K.~Lee, 
Phys.\ Lett.\ B {\bf 495}, 231 (2000) 
[hep-th/0007107]. 
 
 
\bibitem{Lee} 
B.~Lee, K.~Lee and H.~S.~Yang, 
Phys.\ Lett.\ B {\bf 498}, 277 (2001) 
[hep-th/0007140]. 
  
 
  
  
  
\bibitem{Bak} 
D.~Bak, 
Phys.\ Lett.\ B {\bf 495}, 251 (2000) 
[hep-th/0008204]. 
 
 
\bibitem{Aganagic} 
M.~Aganagic, R.~Gopakumar, S.~Minwalla and A.~Strominger, 
``Unstable solitons in noncommutative gauge theory,'' 
hep-th/0009142. 
 
 

 
\bibitem{Harvey} 
J.~A.~Harvey, P.~Kraus and F.~Larsen, 
JHEP{\bf 0012}, 024 (2000) 
[hep-th/0010060]. 
 
 






\bibitem{Gross2} 
D.~J.~Gross and N.~A.~Nekrasov, 
``Solitons in noncommutative gauge theory,'' 
hep-th/0010090. 
    

\bibitem{Hamanaka}
M.~Hamanaka and S.~Terashima,
``On exact noncommutative BPS solitons,''
hep-th/0010221.

\bibitem{Hashimoto2}
K.~Hashimoto,
JHEP{\bf 0012}, 023 (2000)
[hep-th/0010251].


\bibitem{Park} 
D.~Bak, K.~Lee and J.~H.~Park, 
``Noncommutative vortex solitons,'' 
hep-th/0011099. 
 
\bibitem{Lozano1} 
G.~S.~Lozano, E.~F.~Moreno and F.~A.~Schaposnik, 
``Nielsen-Olesen vortices in noncommutative space,'' 
hep-th/0011205. 
 
  


\bibitem{Witten}
M.~Mihailescu, I.~Y.~Park and T.~A.~Tran,
``D-branes as solitons of an N = 1, D = 10 non-commutative gauge theory,''
hep-th/0011079;
E.~Witten,
``BPS bound states of D0-D6 and D0-D8 systems in a B-field,''
hep-th/0012054.

\bibitem{Corley}
S.~Corley and S.~Ramgoolam,
``Projector equivalences in K theory 
and families of non-commutative  solitons,''
hep-th/0012217.


 
\bibitem{Hong} 
J.~Hong, Y.~Kim and P.~Y.~Pac, 
Phys.\ Rev.\ Lett.\ {\bf 64}, 2230 (1990); 
R.~Jackiw and E.~J.~Weinberg, 
Phys.\ Rev.\ Lett.\ {\bf 64}, 2234 (1990). 
 
\bibitem{Jackiw2} 
R.~Jackiw and S.~Y.~Pi, 
Phys.\ Rev.\ Lett.\ {\bf 64}, 2969 (1990);
R.~Jackiw and S.~Pi,
Phys.\ Rev.\ D {\bf 42}, 3500 (1990).
 
 
\bibitem{Dunne} 
G.~V.~Dunne, 
``Aspects of Chern-Simons theory,'' 
hep-th/9902115. 
    

\bibitem{Chu} 
C.~Chu, 
Nucl.\ Phys.\ B {\bf 580}, 352 (2000) 
[hep-th/0003007]. 
 
\bibitem{Bichl} 
A.~A.~Bichl, J.~M.~Grimstrup, V.~Putz and M.~Schweda, 
JHEP{\bf 0007}, 046 (2000) 
[hep-th/0004071]. 
 
 
\bibitem{Chen} 
G.~Chen and Y.~Wu, 
Nucl.\ Phys.\ B {\bf 593}, 562 (2001) 
[hep-th/0006114]. 
 
 
\bibitem{Grandi} 
N.~Grandi and G.~A.~Silva, 
``Chern-Simons action in noncommutative space,'' 
hep-th/0010113. 
 

 
\bibitem{Brodie} 
J.~H.~Brodie, L.~Susskind and N.~Toumbas, 
JHEP{\bf 0102}, 003 (2001) 
[hep-th/0010105]; 
S.~S.~Gubser and M.~Rangamani, 
``D-brane dynamics and the quantum Hall effect,'' 
hep-th/0012155. 
 

%
 
\bibitem{Lozano} 
G.~S.~Lozano, E.~F.~Moreno and F.~A.~Schaposnik, 
``Self-dual Chern-Simons solitons in noncommutative space,'' 
hep-th/0012266. 
  

\bibitem{Khare}
A.~Khare and M.~B.~Paranjape,
``Solitons in 2+1 dimensional 
non-commutative Maxwell Chern-Simons Higgs  theories,''
hep-th/0102016.

\bibitem{Jabbari}
M.~M.~Sheikh-Jabbari,
``A note on noncommutative Chern-Simons theories,''
hep-th/0102092.
  
\bibitem{Susskind}  
L.~Susskind, 
``The quantum Hall fluid and non-commutative Chern Simons theory,'' 
hep-th/0101029. 
 




 

  
 
  
 
\bibitem{Gross}
D.~J.~Gross, A.~Hashimoto and N.~Itzhaki,
``Observables of non-commutative gauge theories,''
hep-th/0008075.

 
  

  
\bibitem{Bak3} 
D.~Bak, K.~Lee and J.~H. Park, 
``Comments on noncommutative gauge theories,'' 
hep-th/0011244. 
 

 
\bibitem{Seiberg} 
N.~Seiberg and E.~Witten, 
JHEP{\bf 9909}, 032 (1999) 
[hep-th/9908142]. 



\bibitem{Jackiw1978}
R.~Jackiw,
Phys.\ Rev.\ Lett.\ {\bf 41}, 1635 (1978).



\bibitem{Kim}
Y.~Kim and K.~Lee,
Phys.\ Rev.\ D {\bf 49}, 2041 (1994)
[hep-th/9211035].


\bibitem{Bak6}
D.~Bak and H.~Lee,
Phys.\ Lett.\ B {\bf 432}, 175 (1998)
[hep-th/9706102].

\bibitem{Bak7} 
M.~Chaichian, A.~Demichev, 
P.~Presnajder, M.~M.~Sheikh-Jabbari and A.~Tureanu,
hep-th/0012175;
D.~Bak, S.~K.~Kim, K.~Soh and J.~H.~Yee, 
Phys.\ Rev.\ Lett.\ {\bf 85}, 3087 (2000) 
[hep-th/0005253]; 
Phys.\ Rev.\ D {\bf 63}, 047701 (2001) 
[hep-th/0006087]. 


\bibitem{Barashenkov}
I.~V.~Barashenkov and A.~O.~Harin,
Phys.\ Rev.\ Lett.\ {\bf 72}, 1575 (1994)
[hep-th/9403056].

\bibitem{Zhang}
S.~C.~Zhang, T.~H.~Hansson and S.~Kivelson,
Phys.\ Rev.\ Lett.\ {\bf 62} (1988) 82.
  
 

  

\end{thebibliography}
\end{document}